\documentclass[preprintnumbers,article,amsmath,amssymb,floatfix,10pt,prd,twocolumn,
superscriptaddress,nofootinbib]{revtex4-2}
\usepackage{bm}
\usepackage{amsfonts}
\usepackage{latexsym}
\usepackage[latin1]{inputenc}
\usepackage{graphicx}
\usepackage{amsmath}
\usepackage{palatino}
\usepackage{mathpazo}
\usepackage{textcomp}
\usepackage{float}
\usepackage{booktabs}
\usepackage{dcolumn}
\usepackage{ragged2e}
\usepackage{hyperref}
\hypersetup{colorlinks,citecolor=blue}
\hypersetup{colorlinks=true,linkcolor=blue,filecolor=magenta,    urlcolor=blue}
\usepackage{amsmath}
\usepackage{subfigure}
\usepackage[]{natbib}
\usepackage{xcolor}
\usepackage{orcidlink}
\usepackage{epsfig}
\usepackage{caption}
\usepackage{subcaption}
\usepackage{commath}
\def\jnl@style{\it}
\def\aaref@jnl#1{{\jnl@style#1}}

\def\aaref@jnl#1{{\jnl@style#1}}

\def\aj{\aaref@jnl{AJ}}                   
\def\apj{\aaref@jnl{ApJ}}                 
\def\apjl{\aaref@jnl{ApJ}}                
\def\apjs{\aaref@jnl{ApJS}}               
\def\apss{\aaref@jnl{Ap\&SS}}             
\def\aap{\aaref@jnl{A\&A}}                
\def\aapr{\aaref@jnl{A\&A~Rev.}}          
\def\aaps{\aaref@jnl{A\&AS}}              
\def\mnras{\aaref@jnl{Mon.~Not.~Roy.~Astron.~Soc.}}             
\def\prd{\aaref@jnl{Phys.~Rev.~D}}        
\def\prc{\aaref@jnl{Phys.~Rev.~C}}  
\def\prl{\aaref@jnl{Phys.~Rev.~Lett.}}    
\def\qjras{\aaref@jnl{QJRAS}}             
\def\skytel{\aaref@jnl{S\&T}}             
\def\ssr{\aaref@jnl{Space~Sci.~Rev.}}     
\def\zap{\aaref@jnl{ZAp}}                 
\def\nat{\aaref@jnl{Nature}}              
\def\aplett{\aaref@jnl{Astrophys.~Lett.}} 
\def\apspr{\aaref@jnl{Astrophys.~Space~Phys.~Res.}} 
\def\physrep{\aaref@jnl{Phys.~Rep.}}      
\def\physscr{\aaref@jnl{Phys.~Scr}}       
\def\commat{\aaref@jnl{Comm.~Math.~Phys.}}              
\def\science{\aaref@jnl{Science}}               
\def\cqg{\aaref@jnl{Classical Quant.~Grav.}}            
\def\jpcs{\aaref@jnl{JPCS}}                                     
\def\ijmpd{\aaref@jnl{Int.~J.~Mod.~Phys.~D}}                    
\def\grg{\aaref@jnl{Gen.~Relat.~Gravit.}}               
\def\rpp{\aaref@jnl{Rep.~Prog.~Phys.}}          
\def\npa{\aaref@jnl{Nucl.~Phys.~A}}        
\def\lrr{\aaref@jnl{Living Rev.~Rel.}}                   
\def\jcap{\aaref@jnl{J.~Cosmology Astropart.~Phys.}}    
\def\rmp{\aaref@jnl{Rev.~Mod.~Phys.}}   
\def\epjc{\aaref@jnl{Eur.~Phys.~J.~C}} 
\def\plb{\aaref@jnl{~Phy.~Lett.~B}} 
\def\mpla{\aaref@jnl{Mod.~Phy.~Lett.~A}} 
\def\arxiv{\aaref@jnl{arxiv.org}}

\allowdisplaybreaks[1]

\begin{document}
\color{black}       
\title{Constraints on bulk viscosity in $f(Q,T)$ gravity from H(z)/Pantheon+ data}

\author{M. Koussour\orcidlink{0000-0002-4188-0572}}
\email[Email: ]{pr.mouhssine@gmail.com}
\affiliation{Department of Physics, University of Hassan II Casablanca, Morocco.} 

\author{Abdelghani Errehymy\orcidlink{0000-0002-0253-3578}}
\email[Email: ]{abdelghani.errehymy@gmail.com}
\affiliation{Astrophysics Research Centre, School of Mathematics, Statistics and Computer Science, University of KwaZulu-Natal, Private Bag X54001, Durban 4000, South Africa.}

\author{O. Donmez}
\email[Email: ]{orhan.donmez@aum.edu.kw}
\affiliation{College of Engineering and Technology, American University of the Middle East, Egaila 54200, Kuwait.}

\author{K. Myrzakulov}
\email[Email: ]{krmyrzakulov@gmail.com}
\affiliation{Department of General and Theoretical Physics, \\ L.N. Gumilyov Eurasian National University, Astana 010008, Kazakhstan.}

\author{M. A. Khan}
\email[Email: ]{mskhan@imamu.edu.sa} 
\affiliation{Department of Mathematics and Statistics, College of Science, Imam Mohammad Ibn Saud Islamic University (IMSIU), P.O. Box 65892, Riyadh 11566, Saudi Arabia.}

\author{B. \c{C}il}
\email[Email: ]{batuhancil@halic.edu.tr} 
\affiliation{Graduate School of Engineering and Science, Istanbul University, Istanbul 34134, Turkey.}
\affiliation{Faculty of Sciences and Literature, Department of Mathematics, Hali\c{c} University, Istanbul 34060, Turkey.}

\author{E. G\"udekli}
\email[Email: ]{gudekli@istanbul.edu.tr} 
\affiliation{Department of Physics, Istanbul University, Istanbul 34134, Turkey.}

%

\begin{abstract}

In this study, we investigate the role of bulk viscosity in $f(Q,T)$ gravity in explaining late-time cosmic acceleration. This model, an extension of symmetric teleparallel gravity, introduces viscosity into cosmic matter dynamics for a more realistic representation. Specifically, we consider the linear form of $f (Q, T) =\alpha Q + \beta T$, where $\alpha$ and $\beta$ are free model parameters. To assess the model, we derive its exact solution and use Hubble parameter $H(z)$ data and Pantheon + SNe Ia data for parameter estimation. We employ the $\chi^2$ minimization technique alongside the MCMC random sampling method to determine the best-fit parameters. Then, we analyze the behavior of key cosmological parameters, including the deceleration parameter, bulk viscous matter-dominated universe density, effective pressure, and the effective EoS parameter, accounting for the viscous type fluid. We observe a transition in the deceleration parameter from a positive (decelerating) to a negative (accelerating) phase at transition redshift $z_t$. The matter density shows the expected positive behavior, while the pressure, influenced by viscosity, exhibits negative behavior, indicative of accelerating expansion. Furthermore, we investigate the energy conditions and find that while the NEC and DEC meet positivity criteria, the SEC is violated in the present and future epochs. The $Om(z)$ diagnostic suggests that our model aligns with quintessence behavior. Finally, our $f(Q,T)$ cosmological model, incorporating bulk viscosity effects, provides a compelling explanation for late-time cosmic behavior, consistent with observational data.\\

\end{abstract}

\maketitle

\section{Introduction}

The accelerated expansion of the universe stands as a significant discovery in modern cosmology, supported by various observational studies such as Type Ia supernovae (SNe Ia) \cite{Riess,Perlmutter}, large-scale structure (LSS) \cite{T.Koivisto,S.F.}, baryon acoustic oscillations (BAO) \cite{D.J.,W.J.}, and cosmic microwave background radiation (CMBR) \cite{R.R.,Z.Y.}. Despite these observations, the underlying cause of this late-time acceleration remains a mystery. Numerous models propose the existence of dark energy (DE), comprising approximately 70\% of the universe \cite{Planck/2020}, which may possess the property of accelerating the universe's expansion. The cosmological constant $\Lambda$ within the framework of General Relativity (GR) serves as a representation of DE, manifesting as a fluid characterized by constant energy density and strong negative pressure. However, this model encounters certain challenges, such as the cosmic coincidence problem \cite{dalal/2001}, where the densities of non-relativistic matter and DE are observed to be of the same order at present. Another significant issue associated with the cosmological constant is the cosmological constant problem. This problem arises from the considerable mismatch between the astronomically observed value of $\Lambda$ (as supported by studies such as \cite{Riess, Perlmutter}) and the value predicted by particle physics for the quantum vacuum energy \cite{weinberg/1989}.

In the quest to address the aforementioned cosmological challenges, various dynamical (time-varying) DE models have been proposed in the literature. These models include the Chaplygin gas model \cite{M.C.,A.Y.}, which considers a fluid with an equation of state (EoS) $p=-\frac{A}{\rho}$, k-essence \cite{T.Chiba,C.Arm.}, which involves a scalar field with a non-standard kinetic term, quintessence \cite{Carroll,Y.Fujii}, which posits a scalar field with a slowly varying potential, and decaying vacuum models \cite{xu/2011,tong/2011,freese/1987,abdel-rahman/1992}, where the cosmological constant decays into the matter over time. Modified gravity theories (MGT) have been extensively studied to elucidate the source of cosmic acceleration and to tackle the issues associated with the cosmological constant model. These theories aim to predict late-time cosmic acceleration by modifying the action of GR. Several possibilities arise within various MGT, including $f(R)$ \cite{appleby/2007,amendola/2007,saffari/2008}, $f(G)$ \cite{cognola/2006,li/2007}, $f(R,T)$ \cite{moraes/2016}, and $f(\mathcal{T})$ \cite{ren/2021,nashed/2015,setare/2013} theories, where $R$, $G$, $T$, and $\mathcal{T}$ represent the Ricci scalar, Gauss-Bonnet scalar, energy-momentum scalar, and torsion scalar, respectively. 

In the late 1990s, the concept of non-metricity theory emerged following the introduction of symmetric teleparallel gravity (STG) \cite{Kalay, Nester, Conroy}. This theory, which lacks torsion and curvature, establishes a link between gravity and the non-metricity tensor along with the associated non-metricity scalar. In 2018, Jimenez et al. extended this theory to $f(Q)$ gravity \cite{Jimenez}, where the gravitational field is described exclusively by the non-metricity scalar $Q$. This theory has effectively explained a range of perturbation and observational data, such as SNe Ia, CMB, and BAO, among others \cite{Koivisto11, Soudi, Salzano, Banos}. 
A study suggests that non-metricity $f(Q)$ gravity could pose a challenge to the $\Lambda$CDM model \cite{Anagnostopoulos}. In addition, researchers have investigated the behavior of black holes using $f(Q)$ gravity \cite{Fell}, demonstrating its applicability in the astrophysical field. 
Wormhole solutions \cite{Hassan11} and Casimir wormholes \cite{Ghosh} have been investigated within the framework of $f(Q)$ gravity. Spherically symmetric configurations in $f(Q)$ gravity have been explored in \cite{Zhai}. Further, various other intriguing applications of $f(Q)$ gravity can be found in the literature \cite{Hohmann,Koivisto,Kuhn,Zhao, Wang2,K1,K2,K3,K4,K5}.

Recently, there has been increasing interest in $f(Q,T)$ theories, which investigate the coupling between matter and geometry through functions of the non-metricity scalar $Q$ and the trace of the energy-momentum tensor $T$ in the Lagrangian \cite{Y.Xu}. Despite being a relatively new gravity theory, $f(Q,T)$ gravity shows promise in various cosmological applications, with several studies already exploring its implications. 
The early studies \cite{Y.Xu} initially explored the cosmological implications of $f(Q, T)$ gravity, while subsequent research \cite{K6,K7} focused on investigating the late-time accelerated expansion with observational constraints for this model. In addition, other aspects such as Baryogenesis \cite{Bhattacharjee}, Cosmological inflation \cite{Shiravand}, and Cosmological perturbations \cite{Najera} have been extensively investigated. However, there has been limited attention given to studying the astrophysical implications of this modified gravity. In a work by Tayde et al. \cite{Tayde1}, static spherically symmetric wormhole solutions in $f(Q,T)$ gravity were examined for both linear and non-linear models under various equations of state. Also, \cite{Sneha2} discusses the thin-shell gravastar model in $f(Q, T)$ gravity. Recently, \cite{Bourakadi} analyzed constant-roll and primordial black holes in $f(Q, T)$ gravity. 

In this context, we investigate the role of bulk viscosity in $f(Q, T)$ gravity in explaining late-time cosmic acceleration. During rapid cosmic expansion, the process of returning to thermodynamic equilibrium leads to the emergence of an effective pressure. This effective pressure is a result of the high viscosity within the cosmic fluid \cite{J.R.,H.O.}. There are two primary viscosity coefficients: shear viscosity $\eta$ and bulk viscosity $\zeta$. Shear viscosity pertains to velocity gradients within the fluid and is typically negligible when the universe is assumed to follow a homogeneous and isotropic Friedmann-Lem\^aitre-Robertson-Walker (FLRW) background. However, some cosmological models have been developed that consider shear viscosity, deviating from the FLRW background assumption \cite{bali/1988,bali/1987,deng/1991,huang/1990}. Bulk viscosity, the focus of our discussion, introduces damping effects associated with volumetric straining. We consider a scaling law for the coefficient of bulk viscosity $\zeta$, which reduces the Einstein case to a form proportional to the Hubble parameter. This scaling law has been demonstrated to be highly beneficial. Samanta et al. \cite{samanta/2017} studied Kaluza-Klein bulk viscous fluid cosmological models within the framework of $f(R, T)$ gravity. Their work explored the validity of the second law of thermodynamics in these models, providing light into the relationship between gravity, viscosity, and thermodynamic principles in cosmology. Satish and Venkateswarlu \cite{satish/2016} investigated cosmological models featuring bulk viscous fluids in the framework of the anisotropic Kaluza-Klein universe, employing $f(R, T)$ gravity. For more cosmological models involving bulk viscous fluids, see \cite{beesham/1993,colistete/2007}. In contrast to the approach taken by Arora et al. \cite{Arora/2022}, who studied the $f(Q, T)$ cosmological model with a bulk viscosity coefficient dependent on velocity, described by $\zeta=\zeta_0+\zeta_1H$, where $\zeta_0$ and $\zeta_1$ are two constants, our study focuses on a simplified scenario where the bulk viscosity coefficient $\zeta$ is treated as a free parameter of the model. The authors determined the best-fit values for the viscosity parameters as $\zeta_0=9.67\pm 0.53$ and $\zeta_1=0.049^{+0.023}_{-0.019}$ using a combination of datasets. This combination includes 57 $H(z)$ data points, 6 data points from the BAO dataset, and the newly published Pantheon samples, consisting of 1048 points.

The current paper is structured as follows. In Sec. \ref{section 2}, we examine the formalism of $f(Q, T)$ gravity theory. 
In Sect. \ref{section 3}, we derive the modified Friedmann equations describing the universe dominated by bulk viscous matter. 
In Sec. \ref{section 4}, we consider a functional form for $f(Q, T)$ and derive the expressions for the Hubble parameter and the deceleration parameter. Further, in Sec. \ref{section 5}, we analyze observational data to determine the best-fit values for the parameters using the $H(z)$ datasets comprising 31 points and the Pantheon+ SNe Ia datasets consisting of 1701 samples. Furthermore, we examine the behavior of various cosmological parameters, including the deceleration parameter, matter-energy density, effective pressure, and effective EoS parameter. Moreover, in Secs. \ref{section 6}, \ref{section 7}, and \ref{section 8}, we check the energy conditions and some diagnostic tools. Finally, in Sec. \ref{section 9}, we summarize our findings.
  
\section{$f(Q,T)$ gravity theory}
\label{section 2}

Next, we will explore an extension of the action in $f(Q)$ STG, which is expressed as \cite{Y.Xu}
\begin{equation}
\label{action}
S = \int \left[ \frac{1}{2 \kappa} f(Q, T) + \mathcal{L}_{M} \right] \sqrt{-g}\ \mathrm{d}^4 x,
\end{equation}
where $f(Q, T)$ is an arbitrary function of the non-metricity $Q$ of spacetime, which quantifies the deviation of the metric from being purely metric, and the trace of the energy-momentum tensor $T$, which describes the distribution of matter and energy in spacetime. When $f(Q, T) = -Q$, the equations of GR are regained. In addition, $\mathcal{L}_{M} $ denotes the Lagrangian density of matter, $g\equiv \det g_{\mu \nu}$, and we adopt the convention $\kappa = 8 \pi G = 1$. 

Now, let us define
\begin{eqnarray}\label{eq:Q}
Q \equiv -g^{\mu \nu} \left( L^{\alpha}_{\ \ \beta\mu} L^{\beta}_{\ \ \nu\alpha} - L^{\alpha}_{\ \ \beta\alpha} L^{\beta}_{\ \ \mu \nu} \right),
\end{eqnarray}
where
\begin{eqnarray}\label{eq:disformation}
L^{\alpha}_{\ \ \beta \gamma} \equiv -\frac{1}{2} g^{\alpha \lambda} \left( \nabla_{\gamma}g_{\beta \lambda} + \nabla_{\beta}g_{\lambda\gamma}-\nabla_{\lambda}g_{\beta \gamma} \right).
\end{eqnarray}

Also, we define the trace of the non-metricity tensor as
\begin{eqnarray}
Q_{\alpha}\equiv Q_{\alpha\ \ \ \mu}^{\ \ \mu}, \ \ \ \ \tilde{Q}_{\alpha}\equiv Q^{\mu}_{\ \  \alpha \mu}.
\end{eqnarray}

Then, we present the superpotential of our $f(Q, T)$ gravity, which is defined as
\begin{eqnarray}\label{eq:superpotential}
\hspace{-0.5cm}&& P^{\alpha}_{\ \ \mu\nu}\equiv \frac{1}{4}\bigg[ -Q^{\alpha}_{\ \ \mu \nu}+ 2Q^{\ \ \ \alpha}_{(\mu \ \ \ \nu)} + Q^{\alpha}g_{\mu \nu} - \tilde{Q}^{\alpha}g_{\mu\nu}\nonumber \\
\hspace{-0.5cm}&&- \delta^{\alpha}_{\ \ (\mu} Q _{\nu)} \bigg] = -\frac{1}{2}L^{\alpha}_{\ \ \mu\nu}+ \frac{1}{4}\left( Q^{\alpha} - \tilde{Q}^{\alpha} \right) g_{\mu \nu} - \frac{1}{4} \delta^{\alpha}_{\ \ (\mu} Q_{\nu)}. \nonumber \\
\end{eqnarray}

After extensive calculations, we derive the expression for $Q$ as
\begin{eqnarray}
&&Q=-Q_{\alpha\mu\nu}P^{\alpha\mu\nu}=- \frac{1}{4}\big( -Q^{\alpha\nu\rho}Q_{\alpha\nu\rho}+ 2 Q^{\alpha\nu\rho} Q_{\rho\alpha\nu} \nonumber \\
&&- 2Q^{\rho}\tilde{Q}_{\rho} + Q^{\rho}Q_{\rho}  \big).
\end{eqnarray}

Furthermore, as usual, we specify
\begin{eqnarray}
T_{\mu\nu} &\equiv& -\frac{2}{\sqrt{-g}}\frac{\delta \left( \sqrt{-g}\mathcal{L}_M \right)}{\delta g^{\mu\nu}}, \\
\Theta_{\mu\nu} &\equiv& g^{\alpha\beta}\frac{\delta T_{\alpha\beta}}{\delta g^{\mu\nu}}.
\end{eqnarray}

This implies that the variation of $\delta T=\delta(T_{\mu\nu} g^{\mu\nu})$, which is given by $\delta T=(T_{\mu\nu} + \Theta_{\mu\nu}) \delta g^{\mu\nu}$.

Following this, we vary the action given in Eq. (\ref{action}) with regard to the components of the metric tensor $g^{\mu \nu}$ to obtain the field equations of the $f(Q,T)$ gravity as
\begin{eqnarray}\label{eq:EOMs}
&&-\frac{2}{\sqrt{-g}}\nabla_{\alpha}\left( f_{Q}\sqrt{-g} P^{\alpha}_{\ \ \mu\nu}   \right) - \frac{1}{2}f g_{\mu\nu} + f_{T}\left( T_{\mu\nu} + \Theta_{\mu\nu} \right) \nonumber \\
&&  -f_{Q}\left( P_{\mu\alpha\beta}Q_{\nu}^{\ \ \alpha\beta} -2 Q^{\alpha\beta}_{\ \ \ \ \mu}P_{\alpha\beta\nu} \right) = \kappa T_{\mu\nu},
\end{eqnarray}
where $f_{Q}$ denotes the derivative with respect to $Q$
and $f_{T}$ represents the derivative with respect to $T$. 

Further, the hypermomentum tensor density is defined as
\begin{eqnarray}\label{eq:hypermomentum}
H_{\lambda}^{\ \ \mu\nu}\equiv \frac{\sqrt{-g}}{2\kappa}f_{T}\frac{\delta T}{\delta \hat{\Gamma}^{\lambda}_{\ \ \mu\nu}} +\frac{\delta \sqrt{-g}\mathcal{L}_{M}}{\delta \hat{\Gamma}^{\lambda}_{\ \ \mu\nu}}.
\end{eqnarray}

Therefore, we obtain the field equations after taking the variation of the gravitational action with regard to the connection,
\begin{eqnarray}\label{eq:feqconnection}
\nabla_{\mu}\nabla_{\nu}\bigg( \sqrt{-g}f_{Q} P^{\mu\nu}_{\ \ \ \ \alpha} +\frac{1}{2}\kappa H_{\alpha}^{\ \ \mu \nu} \bigg)=0.
\end{eqnarray}

Moreover, it is important to mention that the $f(Q,T)$ theory's divergence of the matter energy-momentum tensor can be written as,
\begin{eqnarray}
\hspace{-0.5cm}&&\mathcal{D}_{\mu }T_{\ \ \nu }^{\mu } =\frac{1}{f_{T}-\kappa }\Bigg[-\mathcal{D}%
_{\mu }\left( f_{T}\Theta _{\ \ \nu }^{\mu }\right) -\frac{2\kappa }{\sqrt{-g}}%
\nabla _{\alpha }\nabla _{\mu }H_{\nu }^{\ \ \alpha \mu } \nonumber\\
\hspace{-0.5cm}&&+\kappa \nabla _{\mu }\bigg(\frac{1}{\sqrt{-g}}\nabla _{\alpha }H_{\nu }^{\ \
\alpha \mu }\bigg)-2\nabla _{\mu }A_{\ \ \nu }^{\mu }+\frac{1}{2}%
f_{T}\partial _{\nu }T\Bigg] =B_{\nu }. \nonumber \\
\end{eqnarray}

Consequently, in the $f(Q,T)$ gravity, the non-conservation vector $B_{\nu }$ depends on $Q$, $T$, and the thermodynamic properties of the system; thus, the matter energy-momentum tensor is not conserved, $\mathcal{D}_{\mu }T_{\ \ \nu }^{\mu }=B_{\nu }\neq 0$. The non-conservation of the matter energy-momentum tensor implies the presence of an additional force affecting massive test particles, leading to non-geodesic motion. This phenomenon signifies the energy flow into or out of a particular volume within a physical system. Moreover, a non-zero right-hand side of the energy-momentum tensor indicates the existence of transfer processes or particle production within the system \cite{Y.Xu}. It is worth noting that in the absence of $f_T$ terms, the energy-momentum tensor becomes conserved, as seen in the above equation \cite{Jimenez}.

\section{Motion equations in $f(Q,T)$ gravity}
\label{section 3}

We will now explore the cosmological consequences of the $f(Q,T)$ gravity, assuming that the cosmos is governed by the isotropic, homogeneous, and spatially flat FLRW metric, which is described by \cite{ryden/2003}
\begin{eqnarray}
\label{FLRW}
ds^2 = -dt^2 + a^2(t)\delta_{ij}dx^i dx^j,
\end{eqnarray}
where, $a(t)$ is the scale factor of the universe. The non-metricity obtained for metric (\ref{FLRW}) is
\begin{equation}
    Q=6H^2,
\end{equation}
where $H = \frac{\dot{a}}{a}$ is the Hubble parameter, representing the rate of expansion of the universe, with $\dot{a}$ denoting the derivative of the scale factor $a$ with regard to cosmic time $t$.

To construct the two modified Friedmann equations that describe cosmic evolution, we assume the matter content of the cosmos as a fluid with the viscosity effect, characterized by its energy-momentum tensor, which can be expressed as follows
\begin{equation}
    T_{\mu\nu} = (\rho + \bar{p})u_{\mu}u_{\nu} + \bar{p}g_{\mu\nu}
\end{equation}
where the four-velocity components of the fluid are $u^\mu=(1,0,0,0)$, $\rho$ is the energy density, and $\bar{p}=p-3\zeta H$ is the bulk viscous pressure. In this case, the bulk viscosity coefficient is $\zeta > 0$, and the standard pressure is $p$.

The relationship between matter-energy density and the standard pressure is expressed as \cite{Ren/2006}
\begin{equation}\label{2d}
p=(\gamma-1)\rho.
\end{equation}

Here, $\gamma$ is a constant ranging from 0 to 2. Thus, the effective EoS describing the bulk viscous cosmic fluid can be expressed as \cite{brevik/2005,gron/1990,C.E./1940}
\begin{equation}\label{2e}
\bar{p}= (\gamma-1)\rho -3\zeta H
\end{equation}

In the context of homogeneity and spatial isotropy, a cosmic fluid with viscosity exhibits dissipative effects. The presence of viscosity in a cosmic fluid can diminish its ideal fluid properties, contributing negatively to the total pressure. This aspect is discussed in Refs. \cite{odintsov/2020,fabris/2006,meng/2009}. Thus, the modified Friedmann equations describing the universe dominated by bulk viscous matter in $f(Q,T)$ gravity are given by \cite{Y.Xu}
\begin{eqnarray}
\label{F1}
\kappa \rho &=&\frac{f}{2}-6FH^2 - \frac{2\tilde{G}}{1+\tilde{G}}\left( \dot{F}H +F \dot{H} \right), \\
\kappa \bar{p}&=& -\frac{f}{2}+6FH^2+ 2\left( \dot{F}H +F \dot{H} \right).
\label{F2}
\end{eqnarray}

Here, $F\equiv f_{Q}$ and $\kappa \tilde{G}\equiv f_{T}$. Combining the aforementioned equations yields the evolutionary equation for the Hubble parameter $H$ as
\begin{equation}
\label{HE}
\dot{H} + \frac{\dot{F}}{F}H = \frac{4\pi}{F} \left( 1+\tilde{G} \right) \left( \rho + \bar{p} \right).
\end{equation}

\section{Cosmological $f(Q,T)$ model}
\label{section 4}

To solve Eq. (\ref{HE}) for $H$, we require an additional equation. For this purpose, we will consider a specific and noteworthy $f(Q,T)$ model defined by
\begin{equation}
    \label{fQT}
    f(Q,T)=\alpha Q+\beta T,
\end{equation}
where $\alpha$ and $\beta$ are model parameters. This model, which depicts an exponentially expanding cosmos naturally, was initially proposed by Xu et al. \cite{Y.Xu}. Loo et al. \cite{Loo/2023} used this model to study Bianchi type-I cosmology, employing observational datasets such as the Hubble parameter and SNe Ia. By using the linear functional form of $f(Q,T)$ as given in Eq. (\ref{fQT}), the field equations (\ref{F1})-(\ref{F2}) can be simplified to
\begin{eqnarray}
\label{F1}
\rho &=&\frac{\alpha  \left(\beta  \dot H-3 (\beta +1) H^2\right)}{(\beta +1) (2 \beta +1)}, \\
\bar{p}&=& \frac{\alpha  \dot H}{\beta +1}+\frac{\alpha  \left(\dot H+3 H^2\right)}{2 \beta +1}.
\label{F2}
\end{eqnarray}

Due to the highly non-relativistic nature of matter, we can safely neglect its pressure ($\gamma=1$). The impact of DE on the universe's dynamics is encompassed in the viscous term $-3\zeta H$, which is treated as a pressure-like quantity. Now, using Eq. (\ref{HE}), we derive the following differential equation for our bulk viscous cosmological $f(Q,T)$ model,
\begin{equation}
\label{3e}
    \overset{.}{H}+\frac{3 \left( \beta +1\right) }{ \left( 3\beta
+2\right) }H^{2}+\frac{3\zeta \left( \beta +1\right) \left( 2\beta +1\right) 
}{\alpha \left( 3\beta +2\right) }H=0.
\end{equation} 

Substituting $ \frac{1}{H} \frac{d}{dt}= \frac{d}{dln(a)}$ into Eq. \eqref{3e}, we obtain
\begin{equation}
\label{33e}
    \frac{dH}{dln(a)}+\frac{3 \left( \beta +1\right) }{ \left( 3\beta
+2\right) }H+\frac{3\zeta \left( \beta +1\right) \left( 2\beta +1\right) 
}{\alpha \left( 3\beta +2\right) }=0.
\end{equation} 

By integrating Eq. (\ref{33e}), we derive the expression for the Hubble parameter, given by
\begin{equation}
\label{Hz}
H(z)=\left[ H_{0}+\frac{\zeta \left( 2\beta +1\right) }{\alpha }\right] \left(
1+z\right) ^{\frac{1}{3\beta +2}+1}-\frac{\zeta \left( 2\beta +1\right) }{%
\alpha },
\end{equation}
where $H_0=H(z=0)$ represents the current value of the Hubble parameter. Specifically, for $\alpha=-1$, $\beta=0$, and $\zeta=0$, the solution simplifies to $H(z)=H_0(1+z)^{\frac{3}{2}}$, which describes the standard matter-dominated universe. 

Further, the deceleration parameter is crucial for describing the dynamics of the universe's expansion phase. It is defined as
\begin{equation}
\label{q}
q=-1+\frac{d}{dt}\left( 
\frac{1}{H}\right).
\end{equation}%

By using Eq. (\ref{Hz}) in Eq. (\ref{q}), we have
\begin{equation}
    q(z)=-1+\frac{3 (\beta +1)}{(3 \beta +2) \left[1-\frac{\zeta (2 \beta +1)}{\alpha H_0+\zeta (2 \beta  +1)} (1+z)^{-\frac{3 (\beta +1)}{3 \beta +2}}\right]}.
\end{equation}

\section{Data, approach, and interpreting the results}
\label{section 5}

In this section, we employ various datasets, including observational Hubble data $H(z)$ and Type Ia Supernova (SNe Ia) distance modulus data, to determine the parameter values of our model that accurately describe the different cosmic epochs. To calculate the appropriate values of $H_0$ and the model parameters $\alpha$, $\beta$, and $\zeta$, we include 31 data points from $H(z)$ datasets and 1701 data points from the Pantheon+ samples. We also combine these datasets for a more comprehensive analysis of our viscosity model. To estimate the best-fit values of the parameter space of our viscosity model:
\begin{equation}
\theta_s=(H_0,\alpha,\beta,\zeta),
\end{equation}
we employ Bayesian techniques. This involves using a likelihood function and the Markov Chain Monte Carlo (MCMC) method, which we implement using the Python package \texttt{emcee} \cite{Mackey/2013}. For MCMC sampling, the likelihood function has the following usual exponential form
\begin{equation}
    \mathcal{L}=\exp(-\chi^2/2),
\end{equation}
where $\chi^2$ denotes the chi-squared function. The exact version of the $\chi^2$ function used for each of the datasets is given below. We use the following prior distributions for the free parameters in our investigation:
\begin{equation}
\begin{gathered}
H_0 \in [60,80], \quad \alpha \in[-2,2], \\ 
\beta \in [-2,2], \quad \zeta \in [0,100].
\end{gathered}
\end{equation}

\subsection{$H(z)$ dataset}

The Hubble parameter is a well-known tool for studying cosmic expansion. In redshift terms, it is expressed as
\begin{equation}
  H(z)=\frac{-1}{1+z}\frac{dz}{dt}.
\end{equation}

As $dz$ is obtained from spectroscopic surveys, one can determine the model-independent value of $H(z)$ by measuring $dt$. In this paper, we include 31 data points of $H(z)$ measurements from the Cosmic Chronometers (CC) dataset within the redshift range $0.07 \leq z \leq 2.41$ \cite{Yu_2018,Moresco_2015}. 
The complete list of 31 data points can be found in reference \cite{Loo/2023}. We define the chi-square function to determine the best-fit values of the bulk viscous model parameters $H_0$, $\alpha$, $\beta$, and $\zeta$ as follows, 
\begin{equation}
\chi^2_{H(z)}(\theta_s)=\sum_{i=1}^{N_H}\bigg[\frac{H^\mathrm{th}_i(\theta_s,z_i)-H^\mathrm{obs}_i(z_i)}{\sigma_{H(z_i)}}\bigg]^2,
\end{equation}
where $\sigma_{H(z_{i})}$ indicates the standard deviation of the $i_{th}$ point, $H_{obs}$ and $H_{th}$ represent the actual and predicted values of $H(z)$, respectively, and $N_H$ denotes the total number of data points.

\subsection{Pantheon+ SNe Ia dataset}

In recent years, observational findings related to SNe Ia have confirmed the accelerated expansion phase of our universe. Over the last two decades, there has been a substantial increase in the number of observations of SNe samples. The Pantheon samples are a dataset of 1048 SNe Ia samples that span the redshift range of $0.01 < z < 2.3$. The dataset was released in 2018 \cite{Scolnic/2018}. The observations in this collection come from several low-redshift surveys, the Hubble Space Telescope (HST) surveys, the Pan-STARRS1 Medium and Deep Surveys, the Supernova Legacy Survey (SNLS), and the Sloan Digital Sky Survey (SDSS). Released recently \cite{Scolnic/2022,Brout/2022}, the Pantheon+ sample includes 1701 light curves of $1550$ SNe Ia in the redshift range of $[0.001, 2.26]$. It is assumed that the luminosity distance is \cite{Planck/2020},
\begin{equation}
    D_L(z)=\frac{c(1+z)}{H_0}S_K\bigg(H_0\int^z_0\frac{d\overline{z}}{H(\overline{z})}\bigg),
\end{equation}
where
\begin{equation}
S_K(x)=    \begin{cases}
      \sinh(x\sqrt{\Omega_K})/\Omega_K,\quad ~ \Omega_K >0\\
      x,\quad\quad\quad\quad\quad\quad\quad \quad ~~\Omega_K=0\\
      \sin (x \sqrt{|\Omega_K|})/|\Omega_K|,~~ \Omega_K<0
    \end{cases}\,.
\end{equation}
where $c$ denotes the speed of light. When the universe is spatially flat, we have
\begin{equation}
D_L(z)=c(1+z)\int^z_0\frac{d\overline{z}}{H(\overline{z})}.
\end{equation}

In theory, the distance modulus can be expressed as
\begin{equation}
\mu^{th}=5\log_{10}D_L(z)+\mu_0,\quad \mu_0 = 5 \log_{10} \frac{1}{H_0 Mpc}+25.
\end{equation}

For the Pantheon+ SNe Ia samples, the chi-square function is presented by
\begin{equation}
\chi^2_{SNe}(\theta_s)=\sum_{i,j=1}^{N_{SNe}}\Delta\mu_{i}\left(C^{-1}_{SNe}\right)_{ij}\Delta\mu_{j}, \quad  \Delta\mu_i=\mu^{th}(\theta_s)-\mu_i^{obs},
\end{equation}
where $\mu^{th}$ indicates the distance modulus's expected value based on our model, and $\mu_i^{obs}$ indicates its observed value. By applying the BEAMS with Bias Correction (BBC) approach \citep{kessler2017}, the nuisance parameters in the Tripp formula \citep{tripp1998two} $\mu= m_{B}-M_{B}+\alpha x_{1}-\beta c+ \Delta_{M}+\Delta_{B}$ were found. As a result, the distinction between the absolute magnitude $M_{B}$ $\left(\mu= m_{B}-M_{B}\right)$ and the apparent magnitude $m_{B}$ is now used to define the observed distance modulus. Furthermore, the distance modulus's total uncertainty matrix is represented as
\begin{equation}
C_{SNe} = D_{stat}+C_{sys}.
\end{equation}

Further, we take into account that $D_{ii,stat}=\sigma^2_{\mu(z_i)}$ corresponds to the diagonal matrix of statistical uncertainties. According to Scolnic et al. \cite{Scolnic/2018}, the BBC approach is used to determine systematic uncertainty,
\begin{equation}
C_{ij,sys} = \sum^K_{k=1}\bigg(\frac{\partial \mu^{obs}_i}{\partial S_k}\bigg)
\bigg(\frac{\partial \mu^{obs}_j}{\partial S_k}\bigg)\sigma^2_{S_k},
\end{equation}
where $S_k$ is the systematic error's magnitude, $\sigma_{S_k}$ is its standard deviation uncertainty, and the indexes $\{i,j\}$ designate the distance modulus's redshift bins.

The $\chi^{2}$ function for the sum of the $H(z)$+SNe Ia datasets is now indicated as
\begin{equation}
\chi^{2}_{joint}= \chi^{2}_{H(z)} + \chi^{2}_{SNe Ia}. 
\end{equation}

\subsection{Interpretation of results}

The likelihood contours for the parameters of our bulk viscous cosmological $f(Q, T)$ model are depicted in Fig. \ref{CC+SNe} using the aforementioned data samples. We perform a chi-square minimization for the Hubble parameter $H(z)$ and the Pantheon+ SNe Ia dataset independently, and subsequently for their combination with $H(z)$+SNe Ia. The obtained best-fit values of the model parameters are summarized in Tab. \ref{table} with the 68\% and 95\% confidence levels (CL). Fig. \ref{Hubblecond} provides a comparative analysis between our bulk viscous cosmological $f(Q, T)$ model and the standard $\Lambda$CDM model in cosmology. Here, we consider $\Omega_{m0} = 0.315$ and $H_0 = 67.4 \pm 0.5$ km/s/Mpc for the plot \cite{Planck/2020}. The figure incorporates the Hubble experimental results consisting of 31 data points along with their respective errors, facilitating a clear and detailed comparison between the predictions of our model and the $\Lambda$CDM model. Therefore, as illustrated in Fig. \ref{Hubblecond}, the model exhibits the best fit across past, present, and future epochs.

\begin{table}[h]
\begin{center}
\begin{tabular}{l c c c c}
\hline\hline 
$dataset$              & $H(z)$ & $SNe Ia$ & $H(z)+SNe Ia$ \\
\hline
$H_{0}$ ($km/s/Mpc$) & $67.7_{-1.3}^{+1.3}$  & $67.7^{+1.6}_{-1.5}$  & $67.6^{+1.2}_{-1.1}$ \\

$\alpha$   & $-0.99_{-0.87}^{+0.78}$  & $-1.01_{-0.91}^{+0.82}$  & $-0.73^{+0.65}_{-0.94}$ \\

$\beta$   & $-0.03_{-0.33}^{+0.33}$  & $0.01_{-0.39}^{+0.39}$  & $-0.13^{+0.38}_{-0.30}$ \\

$\zeta$   & $39.2_{-1.9}^{+2.0}$  & $39.3_{-1.9}^{+1.9}$  & $39.3_{-2.0}^{+2.0}$ \\

$q_{0}$   & $-0.31_{-0.3}^{+0.3}$  & $-0.38_{-0.05}^{+0.05}$  & $-0.34^{+0.08}_{-0.08}$ \\

$z_{tr}$   & $0.74_{-0.30}^{+0.31}$  & $1.04_{-0.22}^{+0.1}$  & $0.68^{+0.15}_{-0.23}$ \\

\hline\hline 
\end{tabular}
\caption{The table presents the best-fit model parameters and statistical analyses using the $H(z)$, SNe Ia, and $H(z)$+SNe Ia datasets, along with 68\% and 95\% CL.}
\label{table}
\end{center}
\end{table}

Furthermore, the Hubble constant ($H_0$) is a crucial parameter in cosmology as it determines the rate at which the universe is expanding. Recent measurements of $H_0$ from different cosmological probes have led to tensions, indicating potential inconsistencies in our understanding of the universe's expansion rate (For an in-depth analysis of $H_0$, see Ref. \cite{Valentino/2021}). One of the significant tensions arises from the comparison between the early universe measurements, such as those from the CMB observations by the Planck satellite, and the late-time measurements, such as those from SNe Ia or the local distance ladder. The Planck mission's latest estimate of $H_0$ based on the CMB is around 67.4 km/s/Mpc \cite{Planck/2020}, while the measurements from SNe Ia and other local probes suggest a higher value, around 74 km/s/Mpc \cite{Riess/2019}. In our analysis, the constraint values on the Hubble constant are $H_0=67.7_{-1.3}^{+1.3}$, $H_0=67.7^{+1.6}_{-1.5}$, and $H_0=67.6^{+1.2}_{-1.1}$ with 68\% CL
for $H(z)$, SNe Ia, and $H(z)$+SNe Ia sample, respectively. These values fall within the range of tensions observed in the broader cosmological community. The tension arises from the discrepancy between these lower values and the higher values obtained from the CMB measurements, highlighting the need for further investigation and possibly new physics to reconcile these discrepancies \cite{Yang/2021,Valentino/2021B}.

\begin{widetext}

\begin{figure}[h]
\centering
\includegraphics[scale=0.4]{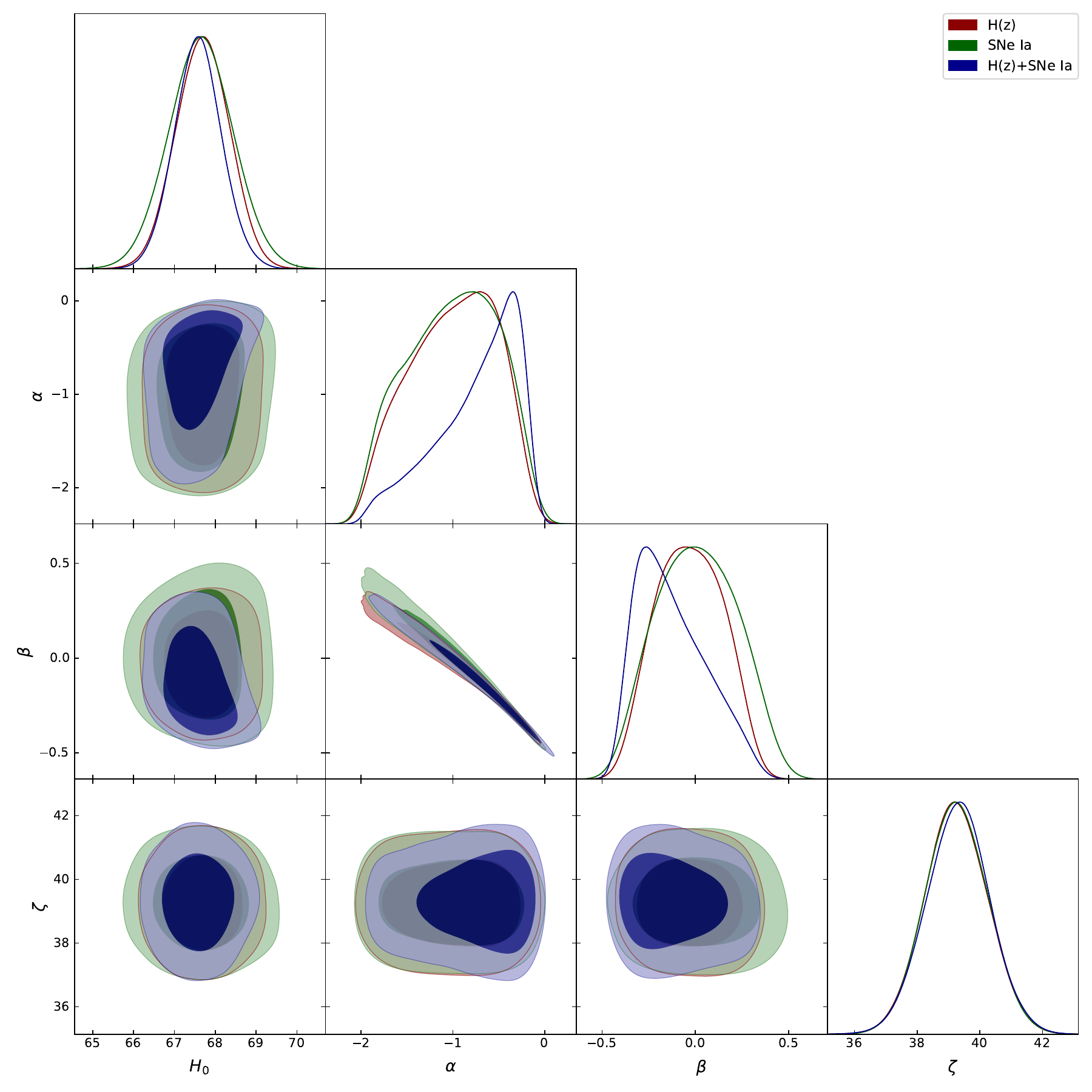}
\caption{The contours representing the $1-\sigma$ and $2-\sigma$ CL for the model parameters $H_0$, $\alpha$, $\beta$, and $\zeta$ are derived from the $H(z)$, SNe Ia, and $H(z)$+SNe Ia datasets.}\label{CC+SNe}
\end{figure}

\begin{figure}[h]
\centering
\includegraphics[width=18cm,height=6cm]{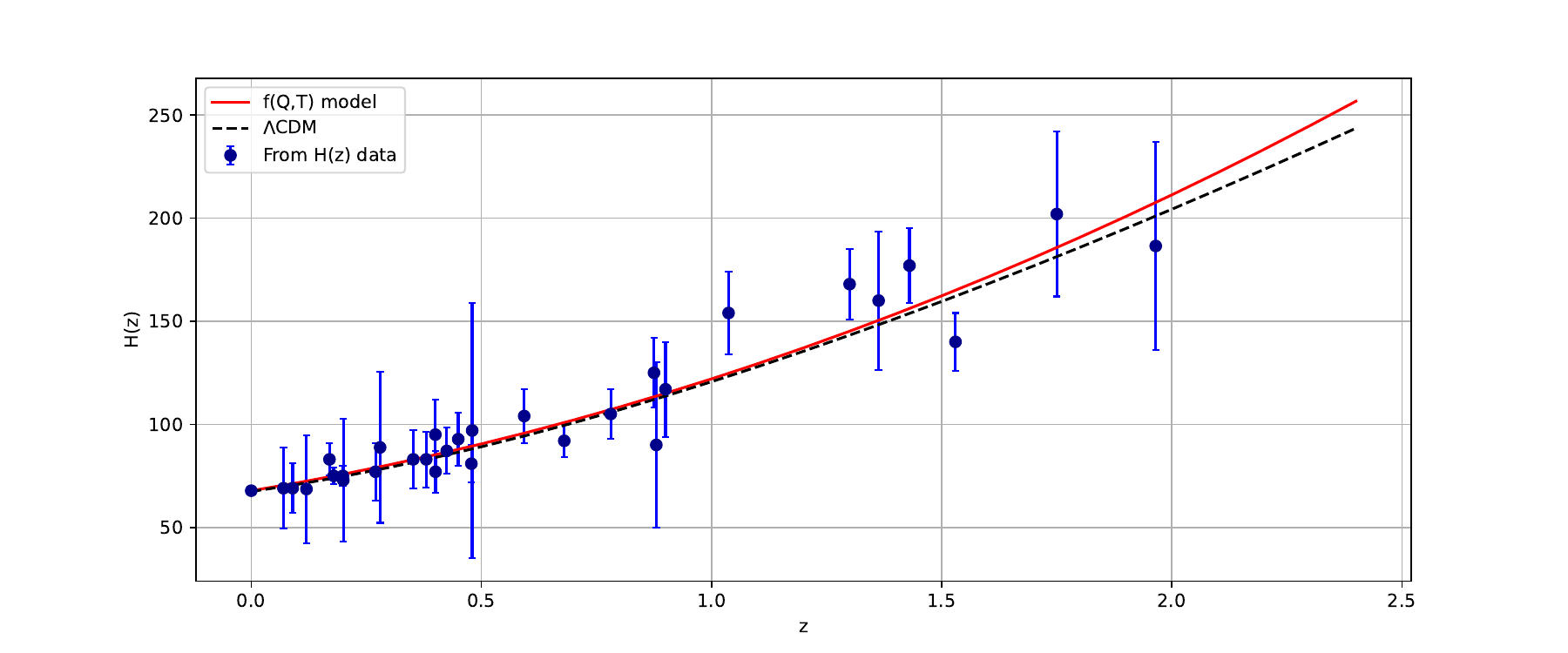}
\caption{The red line represents the evolution of the Hubble parameter for our bulk viscous cosmological $f(Q, T)$ model, using the constrained values of $H_0$, $\alpha$, $\beta$, and $\zeta$. The black dots with blue error bars depict the $H(z)$ data, while the black dotted line shows the evolution of the Hubble parameter for the $\Lambda$CDM model.}\label{Hubblecond}
\end{figure}

\end{widetext}

\begin{figure}[h]
\centering
\includegraphics[scale=0.7]{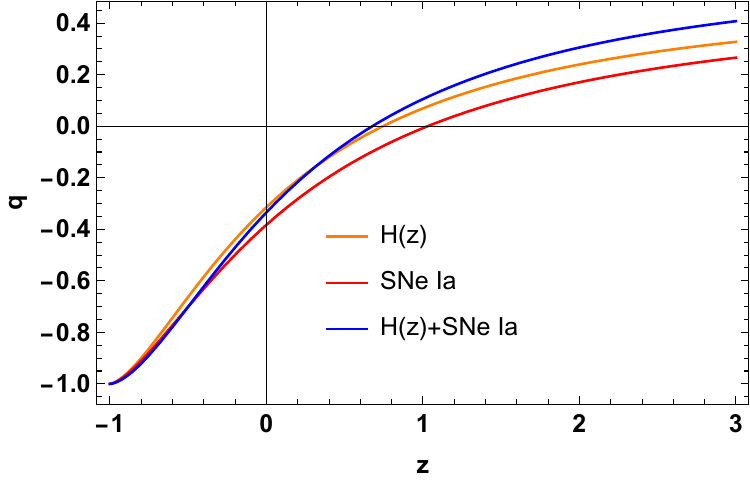}
\caption{The evolution of the deceleration parameter $q$ with respect to the redshift $z$, based on the constrained values of $H_0$, $\alpha$, $\beta$, and $\zeta$, as derived from the $H(z)$, SNe Ia, and $H(z)$+SNe Ia datasets.}\label{F_q}
\end{figure}

\begin{figure}[h]
\centering
\includegraphics[scale=0.7]{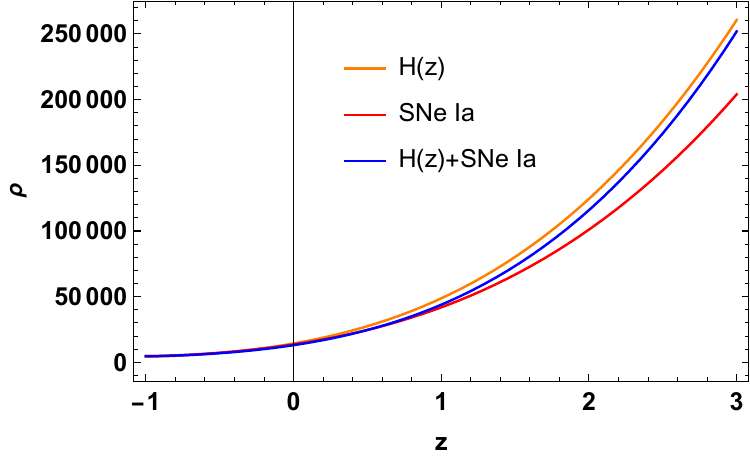}
\caption{The evolution of the energy density $\rho$ with respect to the redshift $z$, based on the constrained values of $H_0$, $\alpha$, $\beta$, and $\zeta$, as derived from the $H(z)$, SNe Ia, and $H(z)$+SNe Ia datasets.}\label{F_rho}
\end{figure}

\begin{figure}[h]
\centering
\includegraphics[scale=0.7]{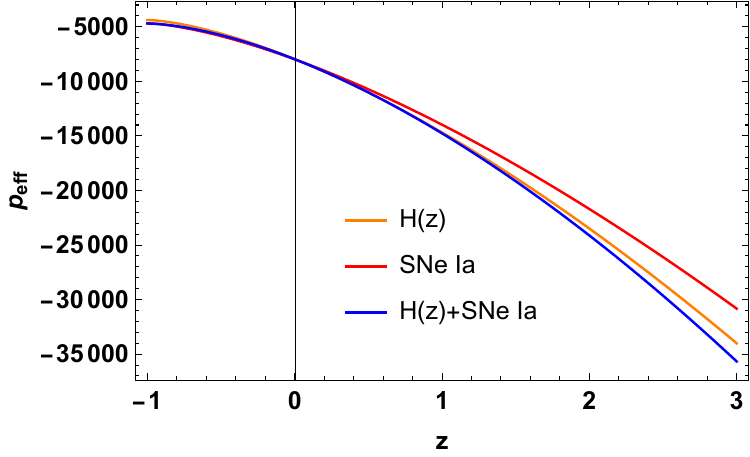}
\caption{The evolution of the effective pressure $p_{eff}$ with respect to the redshift $z$, based on the constrained values of $H_0$, $\alpha$, $\beta$, and $\zeta$, as derived from the $H(z)$, SNe Ia, and $H(z)$+SNe Ia datasets.}\label{F_p}
\end{figure}

\begin{figure}[h]
\centering
\includegraphics[scale=0.7]{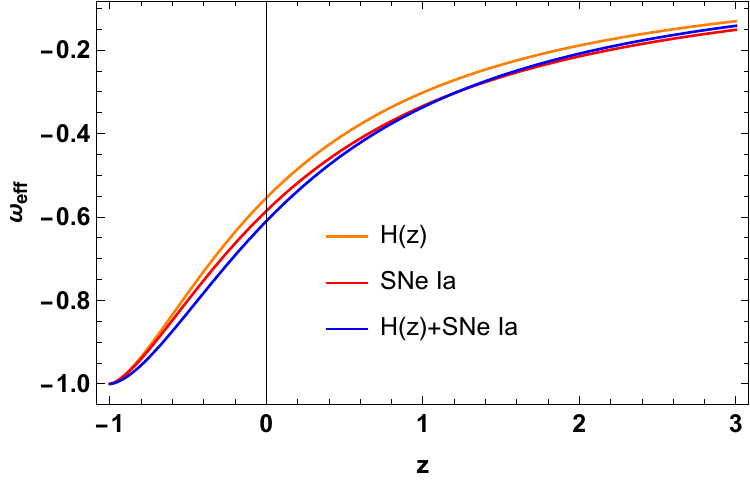}
\caption{The evolution of the effective EoS parameter $\omega_{eff}$ with respect to the redshift $z$, based on the constrained values of $H_0$, $\alpha$, $\beta$, and $\zeta$, as derived from the $H(z)$, SNe Ia, and $H(z)$+SNe Ia datasets.}\label{F_EoS}
\end{figure}

Now, we will discuss the cosmological consequences of the observational constraints we have obtained. For this purpose, we examine the behavior of the deceleration parameter, energy density, the pressure component including viscosity, and the effective EoS parameter. Figs. \ref{F_q}-\ref{F_EoS} depict the evolution of the deceleration parameter ($q$), matter-energy density ($\rho$), effective pressure ($\bar{p}$), and effective EoS parameter ($\omega_{eff}$) as a function of redshift ($z$).

As per cosmological observational data, the acceleration of the universe is a recent development. A comprehensive cosmological model should therefore encompass both the decelerating and accelerating phases of expansion \cite{Mamon/2017}. Consequently, it is crucial to examine the behavior of the deceleration parameter $q$ to fully understand the universe's evolutionary trajectory. The figure illustrates the behavior of $q$ for the corresponding values of model parameters constrained by the $H(z)$, SNe Ia, and $H(z)$+SNe Ia datasets, as shown in Fig. \ref{F_q}. The parameter 
$q$ undergoes a transition from a decelerated phase ($q>0$) to an accelerated phase ($q<0$) at redshift $z_t$. The value of $z_t$ fluctuates within the range of 0.6 to 1.2, consistent with recent observations \cite{Farooq/2013,Cunha/2009,Cunha/2008}. In addition, Tab. \ref{table} displays the current value of the deceleration parameter for the model parameters constrained by the $H(z)$, SNe Ia, and $H(z)$+SNe Ia datasets, aligning with recent studies on the deceleration parameter \cite{Mamon/2017,Gruber/2014}.

Throughout the universe's evolution, the bulk viscous matter-dominated universe density $\rho$ consistently stays positive, increasing with cosmic redshift $z$ (see Fig. \ref{F_rho}). It initiates with a positive value and tends towards zero as $z$ approaches -1. From Fig. \ref{F_p}, the effective pressure component $\bar{p}$ is a decreasing function of redshift, starting from a large negative value and tending towards a small negative value at the present epoch. This negative pressure is attributed to DE in the context of the universe's accelerated expansion. Therefore, the behavior of the effective pressure in our model aligns with this observation, further supporting the consistency of our model with the accelerated expansion scenario. These trends align with the values of model parameters constrained by the $H(z)$, SNe Ia, and $H(z)$+SNe Ia datasets, providing further support for the model's consistency with observational data. 

Moreover, the EoS parameter is instrumental in classifying the different epochs of accelerated and decelerated expansion in the universe. These epochs can be categorized as follows: (a) $\omega=1$ represents a stiff fluid; (b) $\omega=\frac{1}{3}$ depicts the radiation-dominated phase; (c) $\omega=0$ indicates the matter-dominated phase; and (d) $\omega<-\frac{1}{3}$ characterizes the accelerating phase, typically associated with DE domination. Quintessence corresponds to a type of DE with EoS $-1<\omega<-\frac{1}{3}$ that evolves. Phantom energy, on the other hand, has EoS $\omega < -1$ and leads to a Big Rip scenario where the universe expands at an increasing rate, eventually tearing apart all bound structures. Lastly, a cosmological constant, often represented by a constant $\omega = -1$, is a form of DE that does not evolve and is consistent with Einstein's cosmological constant term in the field equations. Now, the effective EoS parameter for our bulk viscous cosmological $f(Q,T)$ model is described by
\begin{equation}
    \omega_{eff}=\frac{\bar{p}}{\rho}=-\frac{3\zeta H}{\rho}.
\end{equation}

The behavior of the effective EoS parameter is presented in Fig. \ref{F_EoS}. In addition, the current values are found to be $\omega_{0}=-0.55$, $\omega_{0}=-0.59$, and $\omega_{0}=-0.61$ for the parameters constrained by the $H(z)$, SNe Ia, and $H(z)$+SNe Ia datasets, respectively \cite{Gruber/2014,EoS1,EoS2,Basilakos}. This alignment indicates that the behavior of the effective EoS parameter depicted in Fig. \ref{F_EoS} confirms the accelerating nature of the universe's expansion phase, resembling quintessence-like behavior.

\section{Energy conditions}
\label{section 6}

Next, we will assess the validity of the obtained solution corresponding to the assumed bulk viscous cosmological $f(Q,T)$ model by applying the energy conditions. These conditions serve as criteria imposed on the energy-momentum tensor to ensure the positivity of energy. They are derived from Raychaudhuri's equation and are defined as follows \cite{Raychaudhuri}:
\begin{itemize}
\item \textbf{Null energy condition (NEC) :} $\rho_{eff}+p_{eff}\geq 0$;  
\item \textbf{Weak energy condition (WEC) :} $\rho_{eff} \geq 0$ and  $\rho_{eff}+p_{eff}\geq 0$; 
\item \textbf{Dominant energy condition (DEC) :} $\rho_{eff} \pm p_{eff}\geq 0$; 
\item \textbf{Strong energy condition (SEC) :} $\rho_{eff}+ 3p_{eff}\geq 0$,
\end{itemize}
where $\rho_{eff}$ represents the effective energy density.

\begin{figure}[h]
\centering
\includegraphics[scale=0.7]{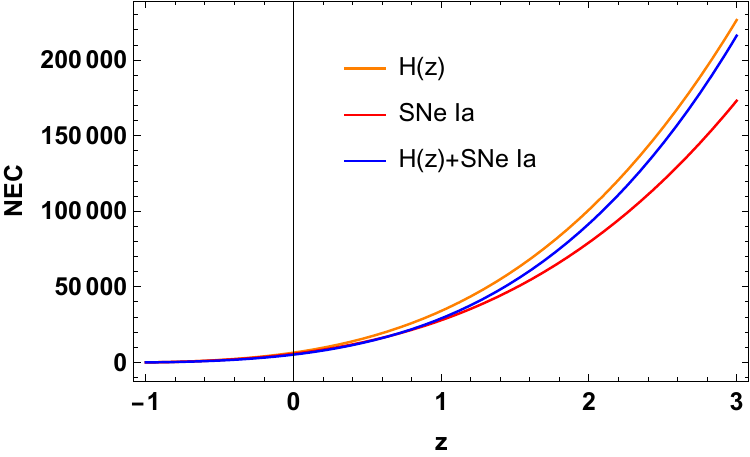}
\caption{The evolution of the NEC $\rho_{eff}+p_{eff}\geq 0$ with respect to the redshift $z$, based on the constrained values of $H_0$, $\alpha$, $\beta$, and $\zeta$, as derived from the $H(z)$, SNe Ia, and $H(z)$+SNe Ia datasets.}\label{F_NEC}
\end{figure}

\begin{figure}[h]
\centering
\includegraphics[scale=0.7]{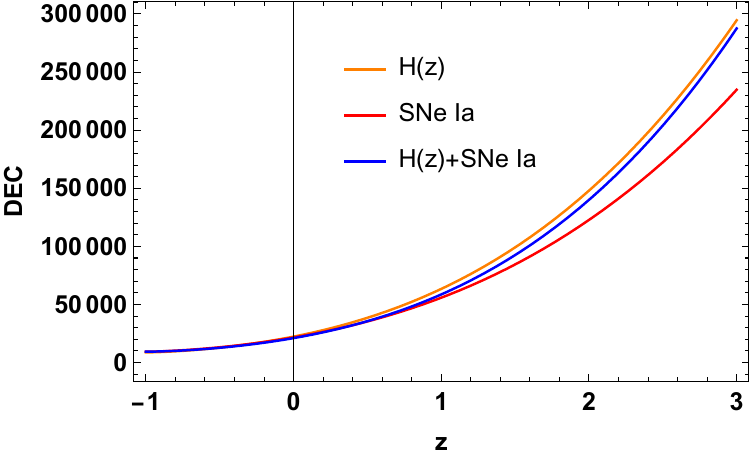}
\caption{The evolution of the DEC $\rho_{eff} \pm p_{eff}\geq 0$ with respect to the redshift $z$, based on the constrained values of $H_0$, $\alpha$, $\beta$, and $\zeta$, as derived from the $H(z)$, SNe Ia, and $H(z)$+SNe Ia datasets.}\label{F_DEC}
\end{figure}

\begin{figure}[h]
\centering
\includegraphics[scale=0.7]{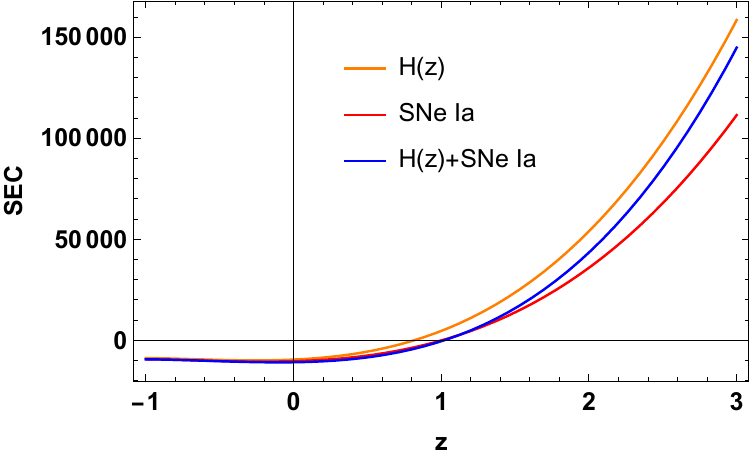}
\caption{The evolution of the SEC $\rho_{eff}+ 3p_{eff}\geq 0$ with respect to the redshift $z$, based on the constrained values of $H_0$, $\alpha$, $\beta$, and $\zeta$, as derived from the $H(z)$, SNe Ia, and $H(z)$+SNe Ia datasets.}\label{F_SEC}
\end{figure}

Observing Figs. \ref{F_NEC} and \ref{F_DEC}, we note that the NEC ($\rho_{eff}+p_{eff}\geq 0$) and DEC ($\rho_{eff} \pm p_{eff}\geq 0$) consistently satisfy the positivity criteria across all redshift values, aligning with the constrained values of the model parameters from observational datasets. Since the WEC ($\rho_{eff} \geq 0$) includes both energy density and NEC, it is also upheld, as indicated by the analysis of Fig. \ref{F_rho}. Moreover, Fig. \ref{F_SEC} illustrates that the SEC ($\rho_{eff}+ 3p_{eff}\geq 0$) is violated in the present and future epochs, indicating a preference for cosmic acceleration \cite{Visser}.

\section{Statefinder diagnostics}
\label{section 7}

As the number of DE models grows, distinguishing between them becomes increasingly challenging. It is crucial to have a sensitive and reliable diagnostic for DE models to differentiate between various cosmological scenarios. Sahni et al. \cite{Sahni/2003} introduced statefinder diagnostics, denoted by the pair $(s, r)$, as new geometrical parameters. These diagnostics serve to characterize different DE models and provide insights into the cosmic acceleration of the universe \cite{Sahni/2003,Alam/2003,Sami/2012,Rani/2015}. The statefinder approach examines the expansion dynamics of the universe through higher derivatives of the scale factor $a(t)$. It offers an alternative viewpoint to the deceleration parameter, which relies on the second derivative of $a(t)$. The statefinder parameters are defined as
\begin{equation}
r=\frac{\overset{...}{a}}{aH^{3}},\text{ \ \ }s=\frac{r-1}{3\left( q-\frac{1%
}{2}\right) }.  
\end{equation}

\begin{figure}[h]
\centering
\includegraphics[scale=0.7]{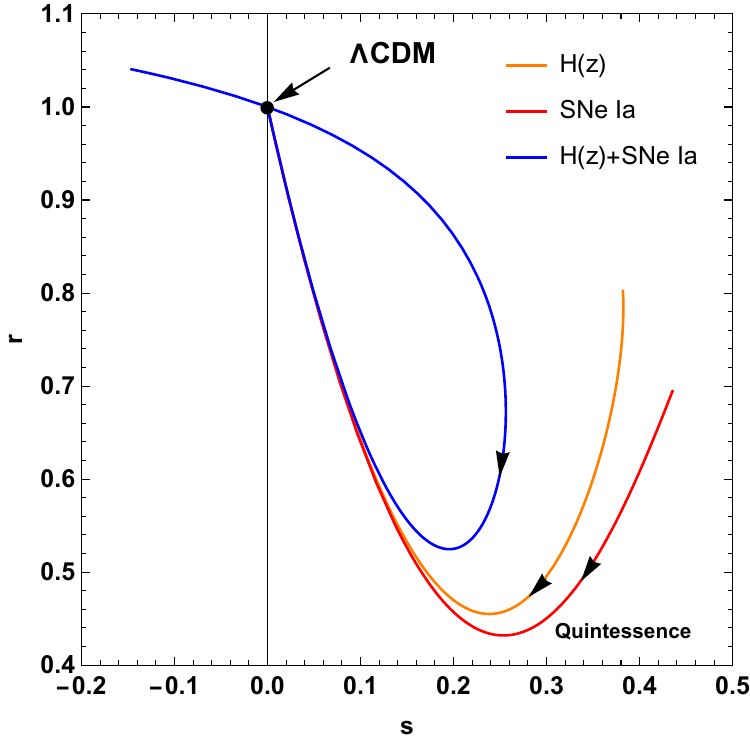}
\caption{The evolution of the statefinder diagnostics $(r, s)$, based on the constrained values of $H_0$, $\alpha$, $\beta$, and $\zeta$, as derived from the $H(z)$, SNe Ia, and $H(z)$+SNe Ia datasets.}\label{F_rs}
\end{figure}

The values $(r > 1, s < 0)$ correspond to the phantom scenario, $(r < 1, s > 0)$ denotes the quintessence-type dark energy, and $(r = 1, s = 0)$ corresponds to the standard $\Lambda$CDM model. The evolution of the statefinder diagnostics $(s, r)$ for the constrained model parameters is illustrated in Fig. \ref{F_rs}. The figure demonstrates the transition of the model from the quintessence region $(r < 1, s > 0)$ to the fixed $\Lambda$CDM point $(r = 1, s = 0)$. In conclusion, our $f(Q,T)$ cosmological model with bulk viscosity is situated in the quintessence region and exhibits behavior similar to a $\Lambda$CDM model in the distant future.

\section{$Om(z)$ diagnostics}
\label{section 8}

The $Om(z)$ diagnostic serves as another valuable tool for categorizing various cosmological models of DE. \cite{Sahni/2008}. It is particularly useful due to its simplicity, relying solely on the first-order derivative of the cosmic scale factor. In the case of a spatially flat universe, the $Om(z)$ diagnostic is expressed as
\begin{equation}
Om(z)= \frac{\big(\frac{H(z)}{H_0}\big)^2-1}{(1+z)^3-1}.
\end{equation}

\begin{figure}[h]
\centering
\includegraphics[scale=0.7]{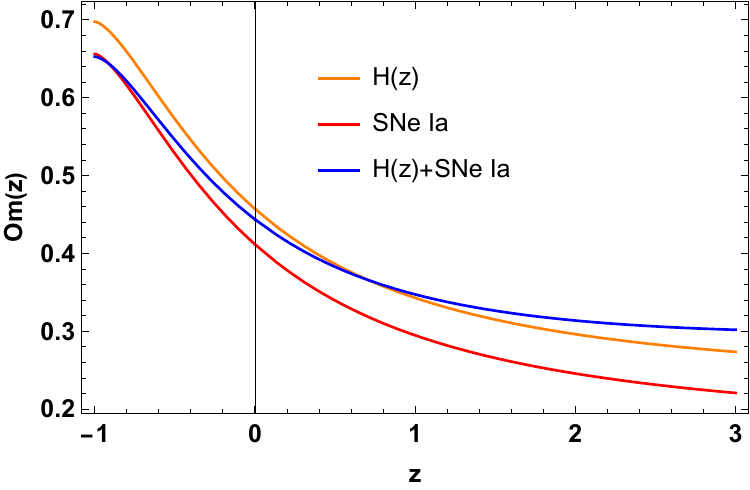}
\caption{The evolution of the $Om(z)$ diagnostic with respect to the redshift $z$, based on the constrained values of $H_0$, $\alpha$, $\beta$, and $\zeta$, as derived from the $H(z)$, SNe Ia, and $H(z)$+SNe Ia datasets.}\label{F_Om}
\end{figure}

The decreasing trend of $Om(z)$ indicates quintessence-like behavior ($\omega_0>-1$), whereas an increasing trend suggests phantom-like behavior ($\omega_0<-1$). A constant $Om(z)$ signifies the $\Lambda$CDM model ($\omega_0=-1$). From Fig. \ref{F_Om}, we observe that the $Om(z)$ diagnostic, calculated for the constrained values of the model parameters, exhibits a decreasing trend across all redshift values. This result from the $Om$ diagnostic test leads us to conclude that our bulk viscous cosmological $f(Q,T)$ model demonstrates quintessence-like behavior, further confirming the behavior of $(r,s)$ and $\omega_{eff}$ in Figs. \ref{F_EoS} and \ref{F_rs}, respectively.

\section{Conclusion}
\label{section 9}

From the viewpoint of hydrodynamics, incorporating a coefficient of viscosity into the cosmic matter content is a natural extension, as the ideal characteristics of a fluid are ultimately abstract concepts. This inclusion allows for a more realistic description of cosmic matter, accounting for its dynamic and dissipative nature in a manner consistent with fluid dynamics. In this article, we examined the role of bulk viscosity in facilitating the late-time acceleration of the universe within the framework of $f(Q,T)$ gravity theory. This theory extends STG \cite{Jimenez}. Xu et al. studied a gravitational action $L$ expressed as an arbitrary function $f(Q,T)$ of the non-metricity $Q$ and the trace of the energy-momentum tensor $T$ \cite{Y.Xu}. Models incorporating geometry-matter coupling bear significant astrophysical and cosmological implications \cite{K6,K7,Bhattacharjee,Shiravand,Najera,Tayde1,Sneha2,Bourakadi}. 

In our analysis, we adopted an $f(Q, T)$ function of the form $f(Q,T) = \alpha Q + \beta T$, where $\alpha$ and $\beta$ are free model parameters. Subsequently, we considered the effective EoS $\bar{p} = p - 3 \zeta H$, which corresponds to the Einstein case value with a proportionality constant $\zeta$ utilized in Einstein's theory. In Sec. \ref{section 4}, we derived the exact solution for the Hubble parameter $H(z)$ by integrating the considered the bulk viscous cosmological $f(Q,T)$ model. In addition, in Sec. \ref{section 5}, we assessed the validity of the considered $f(Q, T)$ model by comparing it with observational datasets, specifically the $H(z)$ datasets and the SNe Ia datasets. The obtained best-fit values are summarized in Tab. \ref{table}, while the $1-\sigma$ and $2-\sigma$ likelihood contours are shown in Fig. \ref{CC+SNe}.

Furthermore, we analyzed the behavior of the deceleration parameter, the bulk viscous
matter-dominated universe density, pressure component incorporating viscosity, and the effective EoS parameter as functions of redshift, as depicted in Figs. \ref{F_q}-\ref{F_EoS}. The deceleration parameter $q(z)$ exhibits a smooth transition from a decelerated to an accelerated phase of expansion, as depicted in Fig. \ref{F_q}. Fig. \ref{F_rho} illustrates that the cosmic matter-energy density is positive and increases with cosmic redshift $z$, reaching $\rho \rightarrow 0$ when $z\rightarrow
-1$. The effective pressure component, presented in Fig. \ref{F_p}, demonstrates negative behavior, indicative of the accelerating expansion of the universe. Further, we determined the current value of the effective EoS parameter is found to be $\omega_{0}=-0.55$, $\omega_{0}=-0.59$, and $\omega_{0}=-0.61$ for the parameters constrained by the $H(z)$, SNe Ia, and $H(z)$+SNe Ia datasets, respectively. Consequently, the evolution of the EoS parameter in Fig. \ref{F_EoS} confirms the accelerating nature of the universe's expansion phase, resembling quintessence-like behavior. Then, the energy conditions depicted in Figs. \ref{F_NEC}-\ref{F_SEC} indicate positivity criteria across all redshift values for the NEC and DEC. However, there is a violation for the SEC, occurring in the present and future epochs, which supports the observed acceleration of the universe. Finally, the $Om(z)$ and statefinder diagnostics, as shown in Figs. \ref{F_Om} and \ref{F_rs}, support the notion that our cosmological model, incorporating bulk viscosity in $f(Q,T)$ gravity, aligns with quintessence-type DE. This finding further validates the behavior of the effective EoS parameter $\omega_{eff}$.

In conclusion, our cosmological $f(Q, T)$ model, which includes the effects of bulk viscosity in the fluid, effectively explains the late-time cosmic behavior of the universe and is consistent with observational data.

\section*{Data Availability Statement}
There are no new data associated with this article.

\section*{Acknowledgments} 
This research is funded by the Science Committee of the Ministry of Science and Higher Education of the Republic of Kazakhstan (Grant No. AP23487178). AE thanks the National Research Foundation of South Africa for the award of a postdoctoral fellowship.

\end{document}